# Heat Transfer in the Flow of a Cold, Two-Dimensional Draining Sheet over a Hot, Horizontal Cylinder

**Jian-Jun SHU** and **Graham WILKS**

School of Mechanical & Aerospace Engineering, Nanyang Technological University
50 Nanyang Avenue, Singapore 639798

## Abstract

An accurate and comprehensive numerical solution to the parabolic free boundary problem arising from thin film flow with both velocity and temperature distribution at large Reynolds numbers is obtained using a modified Keller box method. A detailed numerical solution scheme is presented to establish the flow and heat transfer characteristics of the draining film flow. A solution is obtained for selecting representative values for the basic parameters of heat transfer in the flow of a cold two-dimensional draining sheet over a hot horizontal cylinder. These indicate the underlying features of developing film thickness, velocity, and temperature distribution. Good agreement with the approximations is obtained, providing basic confirmation of the validity of the numerical results presented. Numerical solutions, supplemented with parameter values associated with specified configurations and operating conditions, can be readily invoked to establish the details of engineering practice.

## 1. Introduction

The flow of a thin liquid film over a horizontal cylinder under gravity frequently occurs in a variety of industrial heat transfer applications, such as heat exchange in desalinization apparatus [1], power and process condensers [2], and film cooling [3]. To understand the operation, particularly the efficiency of these processes, it is important to study such processes in detail.

An approximate solution was obtained by Abdelghaffer *et al*. [4] using the Pohlhausen integral momentum technique, assuming an approximate velocity profile across the film thickness. A numerical solution was obtained by Hunt [5] using a modified Keller box method that accommodated the outer free boundary.

In this paper, an accurate and comprehensive numerical solution of both velocity and temperature distributions is obtained. This solution can be applied to a variety of problems [6-9]. Specifically, heat transfer in the flow of a cold two-dimensional draining sheet over a hot horizontal cylinder is investigated.

## 2. Modelling

The problem to be examined concerns film cooling that occurs when a cold vertically draining sheet strikes a hot horizontal cylinder. Although a sheet of fluid draining under gravity is accelerated and thinned upon impact [10,11], it is reasonable to model the associated volume flow as a jet of uniform velocity $U_0$, temperature $T_0$, and semi-thickness $H_0$, as illustrated in Figure 1. The film Reynolds number can be defined as $R_e = \frac{U_0 a}{\nu}$ where $\nu$ is the kinematic viscosity and $a$ is the radius.

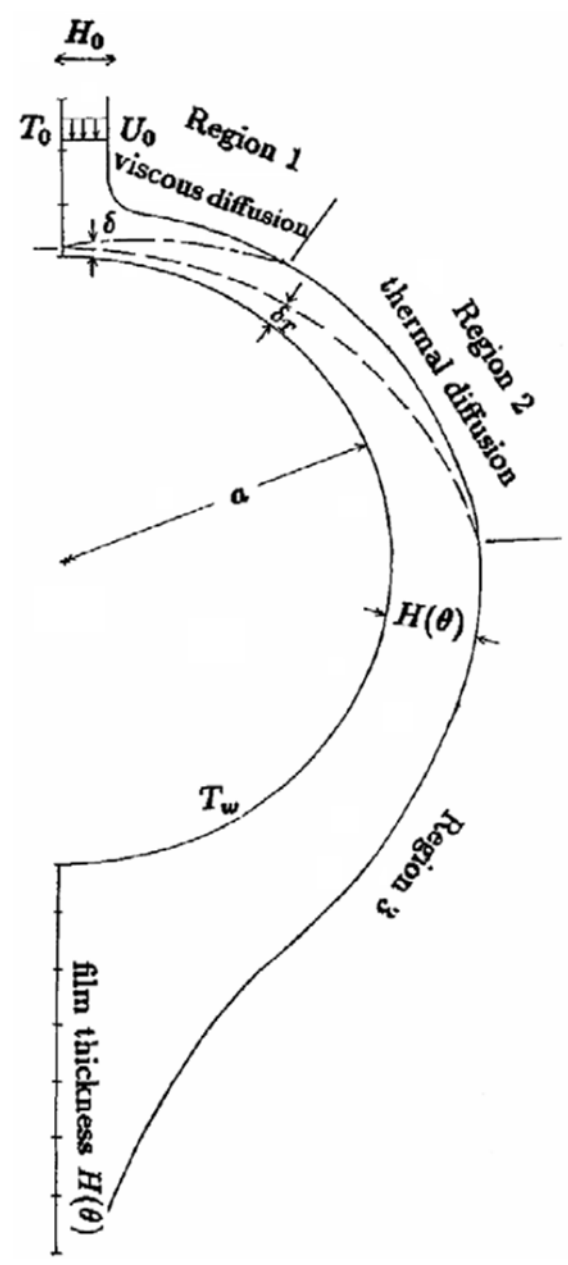

**Fig. 1. Vertical sheet and resultant film for cylinder**

## 3. Governing Equations

The flow under investigation is modelled as a steady two-dimensional flow of an incompressible fluid. In the absence of viscous dissipation, the equations expressing the conservation of mass, momentum, and energy are consequently

$$
\begin{aligned}
&\frac{\partial v_\theta}{\partial \theta}+\frac{\partial}{\partial r}\left(r v_r\right)=0, \\
&\frac{v_\theta}{r}\frac{\partial v_\theta}{\partial \theta}+v_r\frac{\partial v_\theta}{\partial r}+\frac{v_r v_\theta}{r}=g\sin\theta-\frac{1}{\rho r}\frac{dP}{\partial \theta} \\
&\qquad +\nu\left[\frac{1}{r}\frac{\partial}{\partial r}\left(r\frac{\partial v_\theta}{\partial r}\right)+\frac{1}{r^2}\frac{\partial^2 v_\theta}{\partial \theta^2}+\frac{2}{r^2}\frac{\partial v_r}{\partial \theta}-\frac{v_\theta}{r^2}\right], \\
&\frac{v_\theta}{r}\frac{\partial T}{\partial \theta}+v_r\frac{\partial T}{\partial r}=\kappa\left[\frac{1}{r}\frac{\partial}{\partial r}\left(r\frac{\partial T}{\partial r}\right)+\frac{1}{r^2}\frac{\partial^2 T}{\partial \theta^2}\right],
\end{aligned}
\tag{1}
$$

where $\vec{V}=\left(v_\theta, v_r\right)$ are the velocity components associated with the cylindrical coordinates $\left(\theta, r\right)$ measuring the angular displacement at the top of the cylinder and the radial distance from the center of the cylinder, respectively. $\rho$, $\kappa$, $T$, and $P$ are density, thermometric conductivity, temperature, and pressure, respectively. In the specified physical settings, the equations are to be solved subject to the following conditions.

(i) No slip boundary condition on the wall requires

$$v_\theta=v_r=0 \quad \text{on} \quad r=a,\ 0\le\theta\le\pi. \tag{2}$$

(ii) Temperature on the wall is assumed constant

$$T=T_w \quad \text{on} \quad r=a,\ 0\le\theta\le\pi. \tag{3}$$

(iii) On the free surface of the film, the shear stress can be considered negligible, so

$$r\frac{\partial}{\partial r}\left(\frac{v_\theta}{r}\right)+\frac{1}{r}\frac{\partial v_r}{\partial \theta}=0 \quad \text{on} \quad r=a+H\left(\theta\right),\ 0\le\theta\le\pi. \tag{4}$$

(iv) Similarly, in a film cooling environment, such as water surrounded by air, it can be assumed that the heat flux on the free surface is negligible, so

$$\frac{\partial T}{\partial r}=0 \quad \text{on} \quad r=a+H\left(\theta\right),\ 0\le\theta\le\pi. \tag{5}$$

(v) A constraint of volume flow conservation leads to the condition

$$\int_a^{a+H(\theta)} v_\theta\left(\theta, r\right)dr=H_0U_0 \quad \text{for} \quad 0\le\theta\le\pi. \tag{6}$$

## 4. Numerical Solutions

Under the assumption that the film thickness remains thin relative to the characteristic horizontal dimension, the boundary layer treatment of the equations leads to a significant simplification. The following non-dimensional variables and a stream function $\psi$ are introduced

$$
\begin{aligned}
&x=\theta,\quad y=\frac{\sqrt{R_e}\left(r-a\right)}{a},\quad h=\frac{\sqrt{R_e}H\left(\theta\right)}{a}, \\
&\frac{\partial\psi}{\partial x}=-\frac{\sqrt{R_e}v_r}{U_0},\ \frac{\partial\psi}{\partial y}=\frac{v_\theta}{U_0},\quad \varphi=\frac{T-T_w}{T_0-T_w},\quad p=\frac{P}{\rho U_0^2}.
\end{aligned}
\tag{7}
$$

In the limit $R_e\to+\infty$ with $x$ remaining $O\left(1\right)$, and after neglecting the terms of $O\left(\frac{1}{\sqrt{R_e}}\right)$ compared to the unity, the following equations are obtained

$$
\begin{aligned}
&\frac{\partial^3\psi}{\partial y^3}+\frac{1}{F_r}\sin x=\frac{\partial\psi}{\partial y}\frac{\partial^2\psi}{\partial x\partial y}-\frac{\partial\psi}{\partial x}\frac{\partial^2\psi}{\partial y^2}, \\
&\frac{\partial^2\varphi}{\partial y^2}=P_r\left(\frac{\partial\psi}{\partial y}\frac{\partial\varphi}{\partial x}-\frac{\partial\psi}{\partial x}\frac{\partial\varphi}{\partial y}\right),
\end{aligned}
\tag{8}
$$

where $P_r=\frac{\nu}{\kappa}$ and $F_r=\frac{U_0^2}{ag}$ are the Prandtl and Froude numbers, respectively. As in the standard boundary layer theory equation, this means that the pressure across the film remains constant. In the absence of an external pressure gradient and with zero shear assumed on the free surface, the pressure term in (1) is identically zero. Boundary conditions now read

$$
\begin{aligned}
&\psi=0,\ \frac{\partial\psi}{\partial y}=0,\ \varphi=0 \quad \text{on} \quad y=0,\ 0\le x\le\pi, \\
&\psi=\gamma,\ \frac{\partial^2\psi}{\partial y^2}=0,\ \frac{\partial\varphi}{\partial y}=0 \quad \text{on} \quad y=h\left(x\right),\ 0\le x\le\pi, \\
&h=\gamma,\ \psi=y,\ \varphi=1 \quad \text{on} \quad x=0,\ 0<y\le\gamma,
\end{aligned}
\tag{9}
$$

where $\gamma=\frac{\sqrt{R_e}H_0}{a}$. If these equations are used as the basis for the solution, there is a strong contrast between the associated algorithm and that developed later in [12,13]; however, the expectation of an initial Blasius boundary layer within the film can be assimilated into solution scheme by further transformations. Nevertheless, the underlying methodology to the solution remains the same. In each case, discretization aims to generate a simultaneous system of nonlinear equations solved by the Newton iterations. The following coordinate transformation, which simultaneously maps the film thickness onto the unit interval and removes the Blasius singularity at the origin, is introduced

$$x=\xi^2,\qquad y=\frac{\xi\eta\bar{h}}{\xi+1-\eta}.$$

The dependent variables are transformed to

$$f=\frac{\xi+1-\eta}{\xi}\psi,\ u=\frac{\partial\psi}{\partial y},\ v=\frac{\xi}{\xi+1-\eta}\frac{\partial u}{\partial y},\ w=\frac{\xi}{\xi+1-\eta}\frac{\partial\varphi}{\partial y}.$$

Equations (8) and (9) are recast as a first order system

$$
\begin{aligned}
&f_\eta=\frac{\left(1+\xi\right)hu}{\xi+1-\eta}-\frac{f}{\xi+1-\eta},\quad u_\eta=\frac{\left(1+\xi\right)hv}{\xi+1-\eta}, \\
&v_\eta=\frac{v}{\xi+1-\eta}-\frac{\xi^2\left(1+\xi\right)h\sin\xi^2}{\left(\xi+1-\eta\right)^3F_r}-\frac{\left(1-\eta\right)\left(1+\xi\right)hfv}{2\left(\xi+1-\eta\right)^4} \\
&\qquad+\frac{\xi\left(1+\xi\right)h}{2\left(\xi+1-\eta\right)^3}\left(uu_\xi-vf_\xi\right), \\
&\varphi_\eta=\frac{\left(1+\xi\right)hw}{\xi+1-\eta}, \\
&w_\eta=\frac{w}{\xi+1-\eta}-\frac{P_r\left(1-\eta\right)\left(1+\xi\right)hfw}{2\left(\xi+1-\eta\right)^4}+\frac{P_r\xi\left(1+\xi\right)h}{2\left(\xi+1-\eta\right)^3}\left(u\varphi_\xi-wf_\xi\right),
\end{aligned}
\tag{10}
$$

subject to boundary conditions

$$
\begin{aligned}
&f=0,\ u=0,\ \varphi=0 \quad \text{on} \quad \eta=0,\ 0\le\xi\le\sqrt{\pi}, \\
&f=\gamma,\ v=0,\ w=0 \quad \text{on} \quad \eta=1,\ 0\le\xi\le\sqrt{\pi}, \\
&h=\gamma,\ f=f_0\left(\eta\right),\ \varphi=\varphi_0\left(\eta\right) \quad \text{on} \quad \xi=0,\ 0<\eta\le1,
\end{aligned}
\tag{11}
$$

where the initial profiles $f_0\left(\eta\right)$ and $\varphi_0\left(\eta\right)$ are found by putting $\xi=0$ and $h=\gamma$ into (10) and solved subject to the conditions $f=u=\varphi=0$ on $\eta=0$ and $u=\varphi=1$ on $\eta = 1$. This needs to be found numerically. This solution has been successfully tested against previously reported results [14-19].

## 5. Results

In Figure 2, the film thickness distribution around the cylinder is plotted from the numerical solution of $F_r = 1$, $\gamma = 1$ and the individual points calculated by Abdelghaffer *et al.* [4] using approximate theory are included in the graph. The agreement is seen to be remarkably good.

## 6. Conclusion

A detailed numerical solution scheme is presented to establish the flow and heat transfer characteristics of the draining film flow. Solutions are obtained to select the representative values of the basic parameters for the horizontal cylinder case. These indicate the underlying features of developing film thickness, velocity, and temperature distribution. Numerical solutions, supplemented by parameter values associated with specified configuration and operating conditions, can be readily invoked to establish the details of engineering practice.

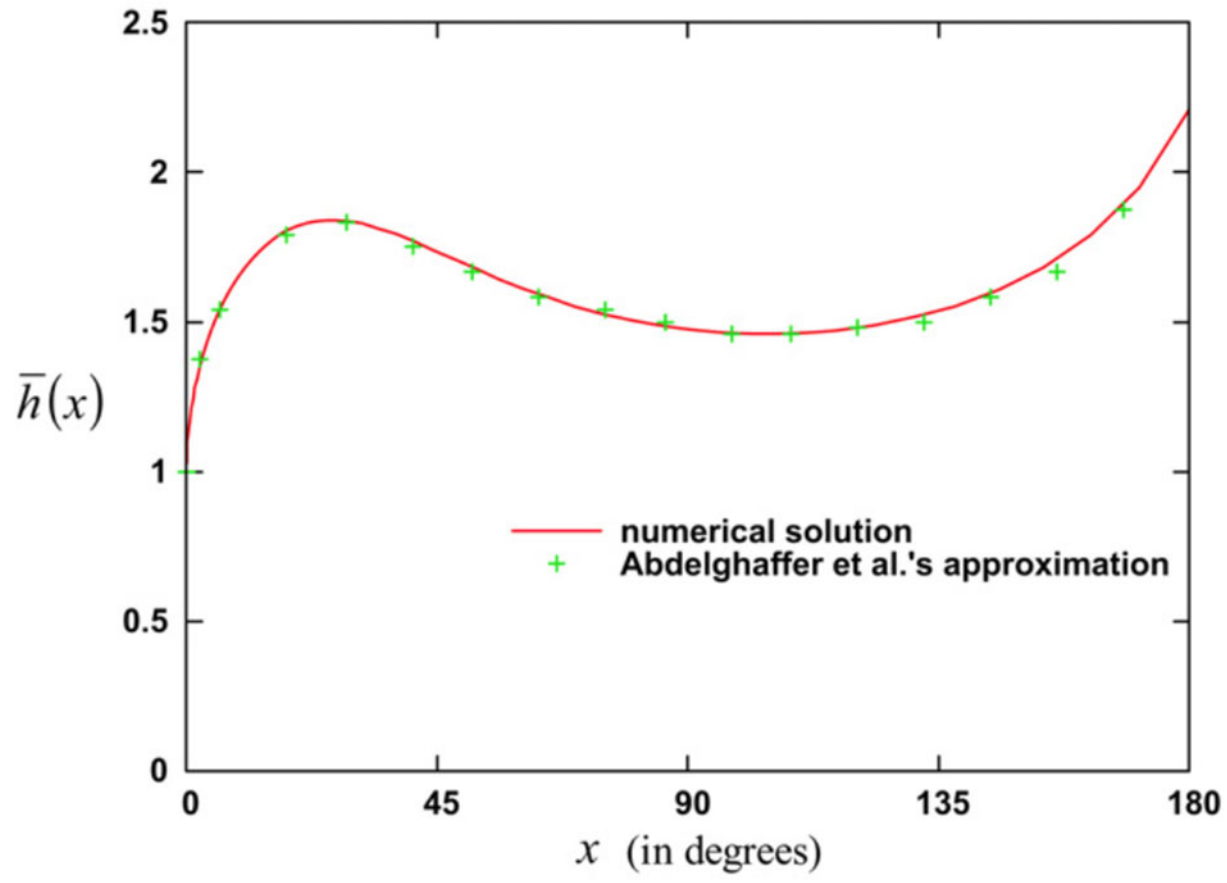

**Fig. 2. Film thickness for numerical solution and approximation at $F_r = 1$ and $\gamma = 1$**